\documentstyle[twoside, epsf]{article}

\input ibvs2.sty

\begin{document}

\IBVShead{NNNN}{15 April 2013}

\IBVStitle{Photometric Analysis of Variable Stars in NGC 299}

\IBVSauth{Sanders~R.J.$^1$; Sarraj~I.$^1$; Schmidtke~P.C.$^1$; Udalski~A.$^2$}

\IBVSinst{School of Earth and Space Exploration, Arizona State University, Box 871404,
Tempe, Arizona 85287 USA}
\vspace{1mm}
\IBVSinsto{E-mail: Raymond.J.Sanders@asu.edu, Ibrahim.Sarraj@asu.edu, Paul.Schmidtke@asu.edu}
\vspace{1mm}
\IBVSinst{Warsaw University Observatory, Aleje Ujazdowskie 4, 00-478 Warsaw, Poland}
\vspace{1mm}
\IBVSinsto{E-mail: Udalski@astrouw.edu.pl}

\IBVSkey{Be stars, X-ray binaries}

\IBVSabs{We have analyzed OGLE-III photometry for stars in the SMC cluster NGC 299.}
\IBVSabs{Two eclipsing binaries and one intrinsic variable (most likely a Be star) are identified.}
\IBVSabs{Unlike other young SMC clusters, no low-amplitude pulsating variables are present.}

\begintext

NGC 299 (RA 00{\hr}53{\mm}24.74{\sec}, DEC $-$72{\deg}11{\arcm}47.6{\arcs},
J2000) is a young and relatively small star cluster in the Small Magellanic Cloud.
It is classified as a Type I cluster on the SWB scale (Searle, Wilkinson, \&
Bagnuolo 1980). Matteucci et al. (2002) analyzed a color-magnitude diagram (CMD)
from $B$ and $V$ observations and, based on the brightness of the upper
main-sequence termination point, estimated the cluster's age to be 15-20 Myr.
They commented on the three brightest stars in the field and suggested these
might be He-burning giants. We further discuss the nature of these stars
later in this paper. Rafelski \& Zaritsky (2005), using data from the Magellanic
Clouds Photometric Survey (MCPS; Zaritsky et al. 2002), calculated integrated
colors for a sample of SMC clusters. From a comparison of these colors with
simple stellar-population models, they derived an extinction-corrected age of
28-100 Myr for NGC 299. Fitting an isochrone model to a MCPS CMD, Glatt, Grebel,
\& Koch (2010) found a cluster age of $\sim$25 Myr. Piatti et al. (2008) fitted a
CMD on the Washington photometric system with an isochrone model to obtain
an age of 25$^{+6}_{-5}$ Myr. They also examined the stellar density profile
(stars per unit area, unweighted by luminosity) and determined the cluster's full
width at half maximum ($r_{\rm FWHM}$) and outer radius ($r_{\rm cls}$) that we
adopt below. Hill \& Zaritsky (2006) showed the brightness profile of NGC 299
is well modeled by a King model.

\vspace{2mm}

A significant percentage of hot stars in the SMC are photometrically variable. This
behavior is most pronounced for Be stars. Diago et al. (2008) analyzed MACHO
observations for a large sample of spectroscopically selected stars and found that 4.9\%
(9 out of 183) of B stars and 25.3\% (32 out of 126) of Be stars were low-amplitude,
short-period pulsating variables. Similarly, in a study of OGLE-II data for NGC 330
(a SMC cluster notable for its large population of Be stars), Schmidtke, Chobanian,
\& Cowley (2008) found that pulsations were present in $>$20\% of their entire sample.
The percentage was even greater for known Be stars. Although no comprehensive
spectroscopic study of NGC 299 has been undertaken, Martayan et al. (2007), in a
survey of SMC B and Be stars, investigated four stars that might be cluster members.
The three brightest of these (i.e., \#11617, \#11998, and \#14323 on the numbering system
discussed below) were found to be evolved B-type stars, while the faintest one (\#11979)
was classified as type B3 IVe. 

\vspace{2mm}

We present here an analysis of 8 seasons of OGLE-III photometry (Udalski 2003;
Udalski et al. 2008) for NGC 299, which lies in field SMC102.1. An $I$-band finding
chart is shown in Fig.\ 1. The two inner circles mark radii corresponding to
$r_{\rm FWHM}$ (=12.6{\arcs}) and $r_{\rm cls}$ (=29.4{\arcs}), respectively. All stars
within the outer circle ($r_{\rm lim}$=45{\arcs}) were examined. The central
portion of the cluster is dominated by light from a very bright star that lies $\sim$1{\arcs}
from the center. Hence, the central region is not well resolved, and the OGLE-III star
closest to the center position is at $r$=5{\arcs}.

\vspace{2mm}

A CMD from $V$ and $I$ data in the OGLE-III photometry maps is shown in Fig.\ 2.
A box is drawn around hot stars on or near the upper main sequence. Stars in this
region of the diagram are likely to show short-period, low-amplitude pulsations
(e.g. see Balona 2010; Ko\l{}aczkowski et al.\ 2006; Moffat 2012). The faint limit of
the box is set at $I$=18.2, which corresponds to the expected brightness of a B5 V star
at the distance of and with an extinction appropriate for the SMC. Within the box there
are 27 stars, of which 6 have $r{>}r_{\rm cls}$. These outliers are unlikely to be cluster
members. When scaled by area on the sky, we estimate that $\sim$1 (out of 12) of the
hot stars with $r{<}r_{\rm FWHM}$ is not a cluster member. Similarly, about 4 stars (out of 9)
with $r_{\rm FWHM}{<}r{<}r_{\rm cls}$ are not members. Based on the young age for
NGC 299, none of the red giants lying to the right side of the box can be a cluster member.

\vspace{2mm}

A plot of $\sigma_I$ vs.\ $I$ from OGLE-III data is shown in Fig.\ 3. Many stars fainter than
$I{\sim}$18.5 have larger than expected photometric dispersion for their mean brightness.
However, almost all faint stars with usually high $\sigma_I$ lie close to the cluster's center
($r{<}r_{\rm FWHM}$), while very few are found outside $r_{\rm cls}$. Hence, we conclude
that it is the lack of consistent spatial resolution of stars near the cluster's core, rather
than photometric variability, that leads to most of the scatter for faint stars in the diagram. 

\vspace{2mm}

All stars within the box shown in Fig.\ 2 were selected for further study. There were $\sim$710
$I$-band observations per star. The time-series data were analyzed for periodic signals
using the Period04 analysis package (Lenz \& Breger 2005). The search covered frequencies
in the range 0-20 day$^{-1}$, which is appropriate for the identification of orbital systems
as well as pulsating B/Be stars. For one star with a decidedly non-sinusoidal light curve
(very narrow eclipses), the phase-dispersion minimization technique of Stellingwerf (1978)
was used to determine the best photometric period.

\vspace{2mm}

As noted by Diago et al.\ (2008), the period analysis of a synoptic data set often shows
significant 1-day aliasing. Many stars in the present sample show this effect, having
fictitious periods comparable to the duration of OGLE-III observing or at high-frequency
aliases near $f$=1, 2, 3, ... day$^{-1}$. The analysis can be further complicated by inadequate
spatial resolution. Only three hot stars in the direction of NGC 299 show a meaningful
photometric signal. Two stars are eclipsing binaries, while a third star shows large, intrinsic
variations that are not periodic. There is no evidence for variability in the four hot stars with
known spectral types. The results are summarized in Table 1 and shown in Fig.\ 4. We note
that no short-period pulsating variables are present in NGC 299. This may be related to age,
as other SMC clusters with a large population of pulsating variables (i.e., NGC 330) are
thought to be slightly older.

\clearpage

\begin{table}[h]
\centerline{{\bf Table~1.} Variable Stars in the Field of NGC 299}
\small
\begin{center}
\begin{tabular}{ccccccccc}
OGLE-III ID & RA & DEC & $I$ & $V{-}I$ & $T_{0}$ & GCVS & $P$ & $r$\\
& (J2000) & (J2000) & (mag) & (mag) & (JD 2453000+) & Type & (days) & (${\prime\prime}$) \\
\hline
\hline
SMC102.1 \#11727 & 0:53:24.14 & -72:11:42.3 & 15.638 & -0.160 & ... & BE? & ... & 6.0\\
SMC102.1 \#11990 & 0:53:25.77 & -72:11:35.3 & 16.812 & -0.084 & 1.022 & EB & 1.59362 & 13.2\\
SMC102.1 \#12553 & 0:53:24.98 & -72:12:14.4 & 17.032 & -0.144 & 1.25 & EA & 14.74086 & 26.9\\
\hline
\end{tabular}
\end{center}

OGLE-III ID refers to the sequence number in field SMC102.1.
$I$ and $V{-}I$ are the mean magnitude and color in the OGLE-III photometry map.
$T_{0}$ represents the time of phase zero, GCVS Type is the variability type,
$P$ denotes the period, and $r$ is the distance from the cluster's center.

\end{table}

Comments on individual sources.

\vspace{2mm}

{\bf SMC102.1 \#11727:}
Based on its brightness (the third brightest of all stars enclosed by the box in Fig.\ 2) and its
proximity to the cluster's center (well within $r_{\rm FWHM}$), this star is likely to be a cluster
member. The photometric variability is typical of a Be star, although no spectrum is available.
The large amplitude (${\Delta}I$=0.35 mag) and long duration ($>$3 yr) of the outburst are
consistent with that of a Be star with a Type-1 (hump-like) light curve (Mennickent et al. 2002).
We searched for low-amplitude pulsations in the first 4 seasons of OGLE-III data (when the light
curve was nearly flat), but found none. An additional search was made of the entire data set,
after subtracting second-order polynomial fits from individual seasons of data. Again, no
low-amplitude pulsations were found. 

\vspace{2mm}

{\bf SMC102.1 \#11990:}
This star lies just outside $r_{\rm FWHM}$ and is likely to be a cluster member. An orbital period
of 1.59 days was found in this $\beta$ Lyrae-type eclipsing binary system. The primary eclipse
has a depth of ${\Delta}I$=0.12 mag, with the depth of secondary eclipse being 0.09 mag. 

\vspace{2mm}

{\bf SMC102.1 \#12553:}
The cluster membership of this star is questionable, as it lies close to $r_{\rm cls}$.
An orbital period of 14.74 days was found in this Algol-type eclipsing binary system.
The light curve shows two very narrow eclipses, neither of which is fully resolved in OGLE-III data.
We tentatively identify the broader eclipse (with a duration of $\sim$0.04P) as the primary and the
narrower eclipse ($\sim$0.02P) as the secondary. Further observations are needed to confirm
these assignments. The depths and durations of the eclipses are consistent with two nearly
identical mid-B main-sequence stars in a 14-day orbit and viewed at an inclination close to 90\deg.
Secondary eclipse falls at phase 0.533, implying an eccentric orbit. 

\vspace{2mm}

The brightest stars in the direction of NGC 299 (labeled A, B, and C in Fig.\ 1) have
been observed by the Two Micron All Sky Survey (2MASS; Cutri et al. 2003). The $K_s$
magnitudes and $J{-}K_s$ colors for these stars are consistent with those of red supergiants in
the SMC (Boyer et al. 2011). All three stars are likely to be cluster members. Stars A and B
are well within $r_{\rm FWHM}$, while star C lies just outside of it. No OGLE-III photometry is
available for these stars. Therefore, we could not examine their long-term variability.

\clearpage

\IBVSfig{10.0cm}{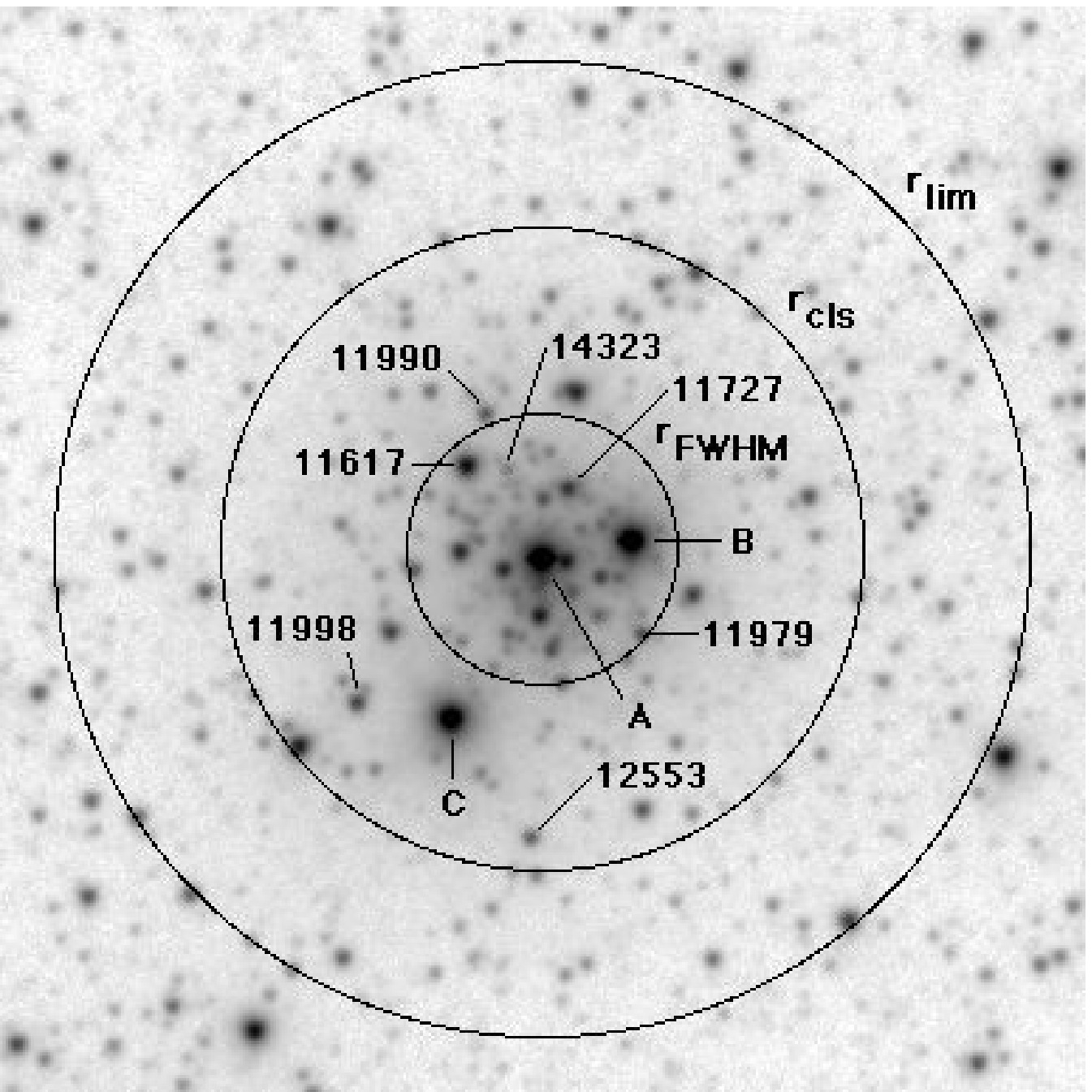}{An $I$-band finding chart for NGC 299. The field of view
is 100{\arcs}$\times$100{\arcs}, with north up and east to the left. Stars with numerical
identifications refer to their sequence number in OGLE-III field SMC102.1, while the
stars labeled A, B, and C are too bright to be in the OGLE data base.}
\IBVSfigKey{NNNN-f1.eps}{NGC 299}{Finding Chart}

\clearpage

\IBVSfig{10.0cm}{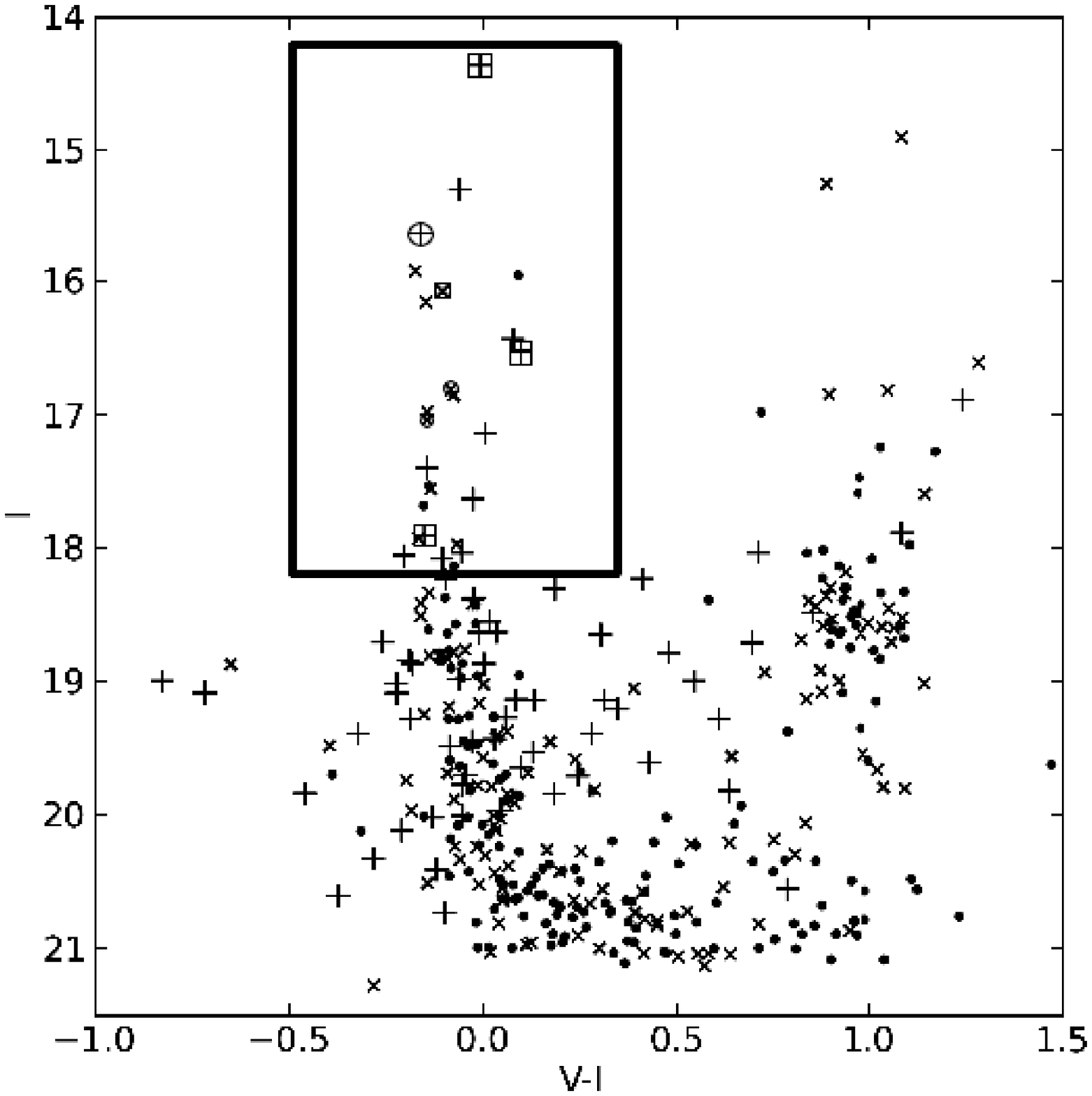}{$I$ vs.\ $V{-}I$ CMD from OGLE-III data for NGC 299.
The box outlines that portion of the diagram in which hot, pulsating variables are likely
to be found. Different symbols indicate relative distances from the cluster's center:
plusses ($+$) for $r{<}r_{\rm FWHM}$,
crosses ($\times$) for $r_{\rm FWHM}{<}r{<}r_{\rm cls}$,
and filled circles ($\bullet$) for $r_{\rm cls}{<}r{<}r_{\rm lim}$.
Open circles ($\bigcirc$) are drawn around variable stars,
while open squares ($\Box$) mark those stars with known spectral types (either B or Be).}
\IBVSfigKey{NNNN-f2.eps}{NGC 299}{CMD}

\clearpage

\IBVSfig{10.0cm}{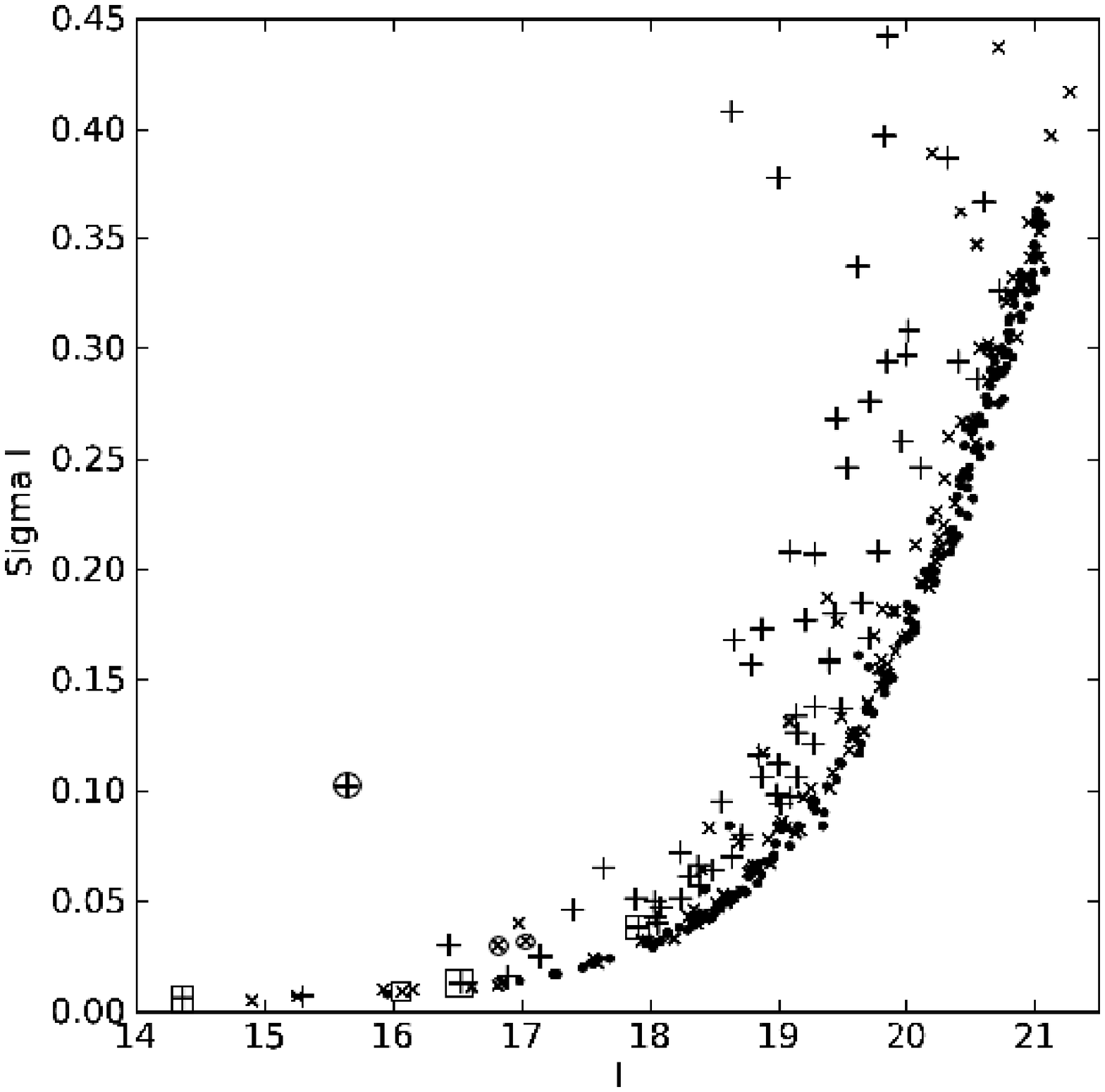}{$\sigma_I$ vs.\ $I$ from OGLE-III data for NGC 299.
Different symbols indicate relative distances from the cluster's center:
plusses ($+$) for $r{<}r_{\rm FWHM}$,
crosses ($\times$) for $r_{\rm FWHM}{<}r{<}r_{\rm cls}$,
and filled circles ($\bullet$) for $r_{\rm cls}{<}r{<}r_{\rm lim}$.
Open circles ($\bigcirc$) are drawn around variable stars,
while open squares ($\Box$) mark those stars with known spectral types (either B or Be).}
\IBVSfigKey{NNNN-f3.eps}{NGC 299}{Dispersion Plot} 

\clearpage

\IBVSfig{18.0cm}{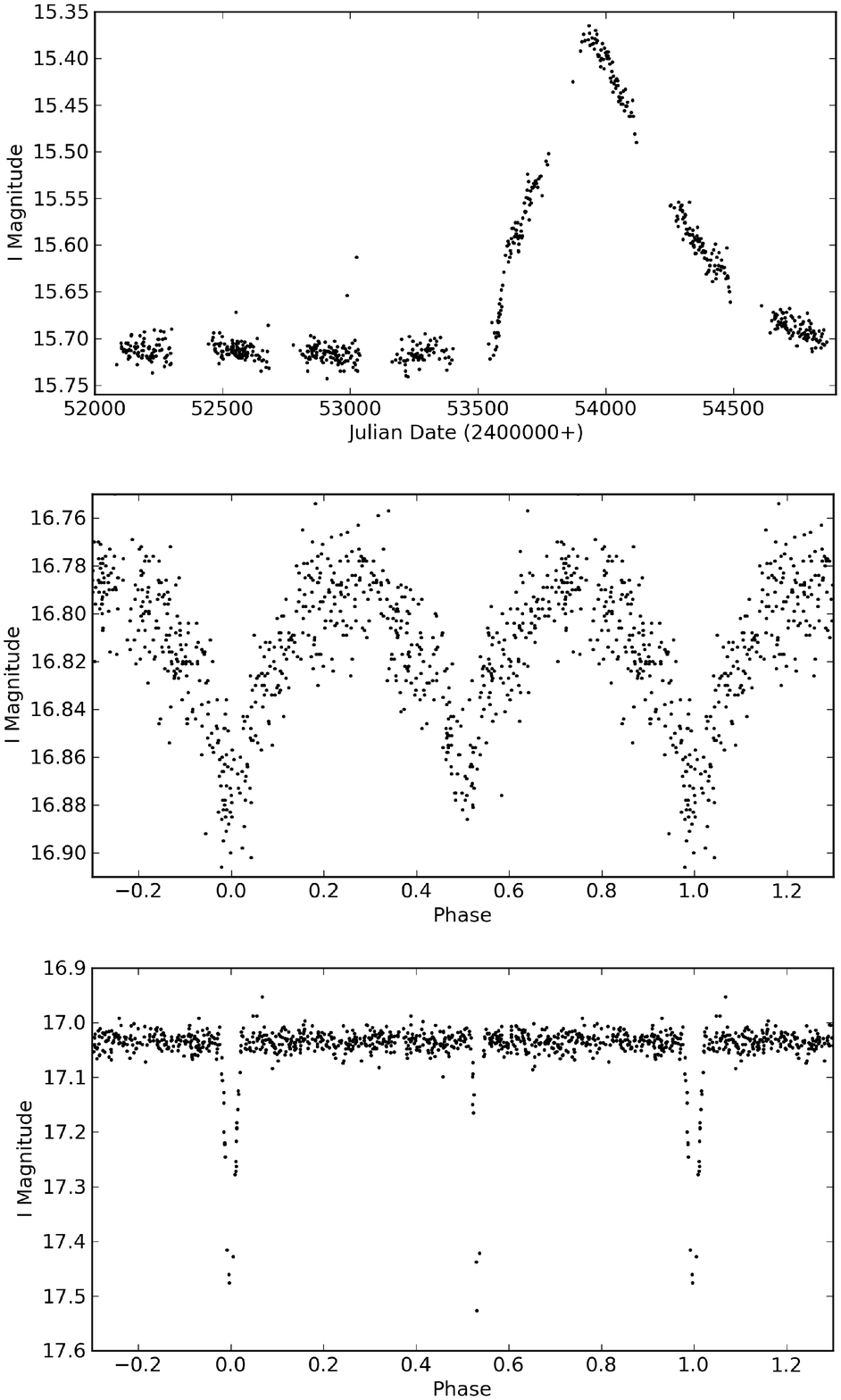}{OGLE-III $I$-band light curves for variable stars in the field
of NGC 299: SMC102.1 \#11727 (top), SMC102.1 \#11990 (middle), and SMC102.1 \#12553
(bottom). See the text and Table 1 for additional information and comments on individual
sources.}
\IBVSfigKey{NNNN-f4.eps}{NGC 299}{Light Curves}

\clearpage

Acknowledgments: The OGLE project has received funding from the European
Research Council under the European Community's Seventh Framework Programme
(FP7/2007-2013) / ERC grant agreement no. 246678 to AU.

\references

Balona, L. A. 2010, Challenges in Stellar Pulsation, Bentham Science Publishers, p.\ 25

Boyer, M. L., et al. 2011, AJ, 142, 103

Cutri, R. M., et al. 2003, VizieR Online Data Catalog, II/246

Diago, P. D., Guti\'{e}rrez-Soto, J., Fabregat, J., \& Martayan, C. 2008, A\&A, 480, 179

Glatt, K., Grebel, E. K., \& Koch A. 2010, A\&A, 517, A50

Hill, A., \& Zaritsky, S. 2006, ApJ, 131, 414

Ko\l{}aczkowski, Z., et al. 2006, Mem.\ Soc.\ Astron.\ Ital., 77, 336 

Lenz, P, Breger, M. 2005, CoAst, 146, 53
 
Matteucci, A., Ripepi, V., Brocato, E., \& Castellani, V. 2002, A\&A, 387, 861

Martayan, C., Fr\'{e}mat, Y., Hubert, A.-M., Floquet, M., Zorec, J., \& Neiner, C. 2007, A\&A, 462, 683

Mennickent, R.E., Pietrzy\'{n}ski, G., Gieren, W., \& Szewczyk, O. 2002, A\&A, 393, 887

Moffat, A. F. J. 2010, in Four Decades of Research on Massive Stars, ASP Conf.\ Ser., 465, 3

Piatti, A. E., Geisler, D., Sarajedini, A., Gallart, C., \& Wischnjewsky, M. 2008, MNRAS, 389, 429

Rafelski, M., \& Zaritsky, D. 2005, AJ, 129, 2701

Schmidtke, P. C., Cobanian, J. B., \& Cowley, A. P. 2008, AJ, 135, 1350

Searle, L., Wilkinson, A., \& Bagnuolo, W. G. 1980, ApJ, 239, 803

Stellingwerf, R.F. 1978, ApJ, 224, 953

Udalski, A. 2003, Acta Astron., 53, 291

Udalski, A., et al. 2008, Acta Astron., 58, 329

Zaritsky, D., Harris, J, Thompson, I. B., Grebel, E. K., \& Massey, P. 2002, AJ, 123, 855

\endreferences

\end{document}